\documentclass[preprint,12pt,3p,times]{elsarticle}
\usepackage{amssymb}
\usepackage{framed}
\usepackage{url}
\journal{Computer Physics Communications}

\begin{document}

\begin{frontmatter}



\title{Strong scaling of general-purpose molecular dynamics simulations on GPUs}


%\author[michigan]{Jens Glaser\corref{cor1}}
%\ead{jsglaser@umich.edu}
\author[michigan]{Jens Glaser}
\author[oakridge]{Trung Dac Nguyen}
\author[michigan]{Joshua A. Anderson}
\author[mellanox]{Pak Lui}
\author[cambridge]{Filippo Spiga}
\author[michiganmatsci]{Jaime A. Millan}
%\author[ohio]{Rong Shi}
%\author[ohio]{D.K. Panda}
\author[minnesota]{David C. Morse}
\author[michigan,michiganmatsci]{Sharon C. Glotzer\corref{cor1}}
\ead{sglotzer@umich.edu}

%\cortext[cor1]{Corresponding authors}
\cortext[cor1]{Corresponding author}
\address[michigan]{Department of Chemical Engineering, 2800 Plymouth Rd, University of Michigan, Ann Arbor, MI 48109, USA}
\address[michiganmatsci]{Department of Materials Science and Engineering, 2300 Hayward St, University of Michigan, Ann Arbor, MI 48109, USA}
\address[oakridge]{National Center for Computational Sciences, Oak Ridge National Laboratory, TN 37831, USA}
\address[mellanox]{Mellanox Technologies, Inc., 350 Oakmead Parkway, Sunnyvale, CA 94085, USA}
\address[cambridge]{High Performance Computing Service, University of Cambridge,
 17 Mill Lane, Cambridge, CB2 1RX, UK}
%\address[ohio]{Department of Computer Science and Engineering, Ohio State University, 2015 Neil Avenue, Columbus, OH 43210, USA}
\address[minnesota]{Department of Chemical Engineering and Materials Science,
421 Washington Street SE, Minneapolis, MN 55455, USA}

\begin{abstract}
We describe a highly optimized implementation of MPI domain decomposition in a
GPU-enabled, general-purpose molecular dynamics code, HOOMD-blue (Anderson and
Glotzer, arXiv:1308.5587).  Our approach is inspired by a traditional CPU-based
code, LAMMPS (Plimpton, J. Comp. Phys.  117, 1995), but is implemented within a
code that was designed for execution on GPUs from the start (Anderson et al.,
J.  Comp. Phys. 227, 2008). The software supports short-ranged pair force and
bond force fields and achieves optimal GPU performance using an autotuning
algorithm. We are able to demonstrate equivalent or superior scaling on up to
3,375 GPUs in Lennard-Jones and dissipative particle dynamics (DPD) simulations
of up to 108 million particles.  GPUDirect RDMA capabilities in recent GPU
generations provide better performance in full double precision calculations.
For a representative polymer physics application, HOOMD-blue 1.0 provides an
effective GPU vs. CPU node speed-up of $12.5\times$.
\end{abstract}


\end{frontmatter}

\section{Introduction}
Graphics processing units (GPUs), massively parallel processors with thousands
of compute cores, represent a disruptive technology shift in simulation
hardware.  Molecular dynamics (MD) simulations once requiring tens or hundreds
of CPU cores are now routinely performed on the researcher's desktop
workstation using only a single GPU. All major MD software packages now take
advantage of single-GPU acceleration, and some of them offer multi-GPU
capabilities.  However, as many-GPU clusters and petascale class supercomputers
such as Titan (18,688 GPUs) and Blue Waters (4,200 GPUs) are becoming a
mainstay of scientific computing, making efficient use of these powerful
resources is key to productivity.

HOOMD-blue, which was the first general-purpose molecular dynamics code written
exclusively for NVIDIA GPUs \cite{Anderson2008}, is extended to run efficiently
on thousands of GPUs in parallel. Today, many, if not all major molecular
dynamics codes support GPU acceleration, including AMBER \cite{Grand2012},
LAMMPS \cite{Plimpton1995, Trott2011,Brown2011,Tang2014} GROMACS
\cite{Pall2013}, NAMD \cite{Stone2007,Phillips2008}, CHARMM \cite{Eastman2013},
DL\_POLY \cite{Lysaght}, ACEMD \cite{Harvey2009}, Desmond \cite{Deng2011},
Espresso \cite{Roehm2012}, and Folding@Home \cite{Eastman2013}.  Since these
codes were designed with CPUs in mind, they take advantage of GPUs at different
levels of efficiency.  In most of these codes, only the dominant
compute-intensive part of the algorithm has been ported to the GPU. The
advantage of such an approach is that it offers significant speed-up vs. a
single CPU core without the need to rewrite a legacy application.  However, it
also means that the code does not take maximum advantage of the GPU if the
particle data structures are not GPU-optimized and data needs to be copied back
and forth between the CPU and the GPU.  Codes designed exclusively for the GPU
include Fen-Zi \cite{Ganesan2011a} or HALMD \cite{Colberg2011}, which implement
only a limited feature set.  HOOMD-blue is unique among all these codes.  It
uses completely device-resident data structures, all simulation work is
performed on the GPU so that the CPU merely acts as a driver for the GPU, and
it offers a rich feature set for general-purpose particle-based simulations
\cite{hoomd-url}.

The first HOOMD releases up to 0.11.3 are highly-tuned for single-GPU
performance and do not support multi-GPU runs. Reducing latency is one of the
biggest challenges in developing a code scalable to many GPUs.  Data
transferred between GPUs moves over the PCIexpress bus (PCIe), whose bandwidth
(up to 16 GB/s) and latency (several $\mu$s) is much slower than on-board GPU
memory (250GB/s, $\sim$ 100 ns). Communicating over PCIe adds latency that is
not present in single-GPU runs. In the strong scaling limit of increasing the
number of GPUs $P$ at constant number $N$ of particles, the work $N/P$
performed by each GPU decreases to the point where it is too small to fully
utilize the device.  Finally, whenever possible, communication should be
overlapped with computation to mask latency.

In developing HOOMD-blue 1.0 with MPI spatial domain decomposition, we
addressed these challenges and here we demonstrate scaling on over 3,000 GPUs.
We show that strong scaling speed-ups in excess of 50x are attainable on the
Titan supercomputer, and weak scaling holds over three orders of magnitude in
system size.  We compare HOOMD-blue to three other implementations of molecular
dynamics (LAMMPS-GPU \cite{Brown2011}, LAMMPS USER-MESO \cite{Tang2014} and
LAMMPS-Kokkos \cite{CarterEdwards2014}) on GPUs and show significant advances
in scaling over these implementations. Moreover, we examine the efficiency of
CUDA-aware MPI and demonstrate superior performance of GPUDirect RDMA
\cite{nvidiagpudirectrdma}, in combination with a high-performance MPI
implementation, \mbox{MVAPICH2} 2.1 \cite{Potluri2013}, for some use
cases.

The remainder of the paper is organized as follows. In
Sec.~\ref{sec:applications}, we give an overview of the research areas within
soft condensed matter physics in which we expect the GPU-based domain
decomposition approach to be helpful, followed by a description of the
capabilities of the present HOOMD-blue version in Sec.~\ref{sec:capabilities}.
Section \ref{sec:comm} presents a detailed description of the new communication
algorithm. In Sec.~\ref{sec:opt_strong_scaling} we further discuss how we
optimized the code for strong scaling on many GPUs. We show results for weak
and strong scaling of a Lennard-Jones (LJ) benchmark in
Sec.~\ref{sec:scaling_titan}.  We show scaling data for a polymer
brush system with various short-ranged force fields in Sec.~\ref{sec:tethers}.
Subsequently  we discuss the efficiency of GPUDirect RDMA for strong scaling
(Sec.~\ref{sec:gdr}). In Sec.~\ref{sec:strong_scaling_dpd}, we discuss a
different benchmark system, DPD. We conclude (Sec.~\ref{sec:conclusion}) with an
outlook on future enabling technologies for GPU-based MD.

\section{Scientific applications}
\label{sec:applications}
HOOMD-blue's modular and compartmentalized, yet easy-to-use design allows
customizability and flexibility for a wide range of applications, as exemplified
by the over 80 published works that use this software to date \cite{hoomdweb}.
Previously, the software has been used for single-GPU simulations, and for
weakly coupled parallel tempering simulations.  Here, we report on how its
features have been enabled with MPI spatial domain decomposition, and give
examples of soft condensed matter research where such functionality may be useful,
such as polymer systems and complex fluids.

Simulations of coarse-grained, bulk polymeric liquids are used for detailed
investigations into the thermodynamic properties of model materials. In those
studies, long polymer chains are required to reduce discretization effects.
Large simulation unit cells are necessary because of the scale separation
between a monomer and the polymer coil, and to simulate many such coils (radii
of gyration).  Typically, bulk polymer liquids or melts have a homogeneous
monomer density and often these systems are effectively incompressible. Such
simulations are particularly amenable to the domain decomposition approach
\cite{Glaser2014,Glaser2014a}.  Similarly suited are, e.g., simulations of
polymer solutions \cite{Reith2011a}, micelle aggregation \cite{Levine2011a},
polymer brushes \cite{Reith2011}, wetting/dewetting phenomena
\cite{Lin2014a,Nguyen2014b,Nguyen2014c}, and glassy polymers \cite{Lam2013}.
Heterogeneous systems, such as micelles in implicit solvent using DPD or
Brownian dynamics, are susceptible to load imbalances \cite{Grime2014} and may
map less well onto the domain decomposition approach.

More complex simulation scenarios include anisotropic particles, polymer
nanocomposites such as tethered nanoparticles \cite{Phillips2010}, colloidal
crystals \cite{Wilms2012a}, structure formation from isotropic pair potentials
\cite{Zhang2013c}, actively driven or self-propelled particles
\cite{Nguyen2012}, and biological applications \cite{Perlmutter2013,
Benedetti2014, Kapoor2013}. The emergent structures of interest, such as a
crystal unit cell or dynamical swarms and clusters, are often much larger than
the size of the constituent building blocks or agents, and many such units or
features need to be simulated, requiring large-scale simulations. Because of
the complexity of the interactions, these simulations tend to be more
compute-intensive and should scale well.

\section{Characterization of HOOMD-blue}
\label{sec:capabilities}
HOOMD-blue is a general-purpose code, though it primarily targets soft-matter
and coarse-grained simulations. Many other specialized MD codes are
available for all-atom models.  As a a general-purpose MD
code, HOOMD-blue is highly customizable and comes with many features, most of
which have been enabled in MPI mode. All of HOOMD-blue's capabilities are
available both on the GPU and the CPU, including MPI domain decomposition, and
the CPU can be used when no GPU is available or for testing purposes.  Here we
briefly describe the features that have been ported to the MPI version.
Existing functionality from previous versions has been fully preserved in single
GPU/core runs. Utilizing the MPI functionality does not require any changes to
existing job scripts.

HOOMD-blue is driven by Python simulation scripts. In multi-GPU/core simulations,
multiple Python interpreters on different MPI ranks process the same input
script. The main simulation engine is written in CUDA and C++.  In HOOMD-blue
1.0, full double precision is available per compile-time option. HOOMD-blue
version 1.0 is optimized for Kepler generation GPUs and also runs on the
previous generation, Fermi. HOOMD-blue is optimized for each new generation of
GPUs as it becomes available. Auto-tuning is used to automatically optimize GPU
kernel execution parameters at runtime (see also Sec.~\ref{sec:autotuner}).

We needed to update a majority of the classes in HOOMD-blue to support MPI,
including the file I/O classes, integrators, pair and bond potentials, and
analyzer classes. File I/O-classes include the initialization and saving of
checkpoints through HOOMD-blue's XML file format, and output of simulation
trajectories, using CHARMM's DCD file format. Integrators include NVE (constant
energy), NVT (isothermal- isochoric), NPH (isotenthalpic-isobaric) and NPT
(isothermal-isobaric) ensembles, which implement the Velocity-Verlet or the
symplectic Martyna-Tobias-Klein update equations \cite{Martyna1994,Martyna1996}
for maximum ensemble stability.  Brownian (or Langevin) dynamics and DPD
\cite{Phillips2011} are supported as well.  Supported pair potentials include LJ,
DPD, Gaussian, Morse, Yukawa and tabulated potentials. For molecular
simulations (with covalent bonds), bond potentials, i.e. harmonic, FENE and
tabulated bonds, and also harmonic and tabulated angle and dihedral potentials,
were updated.  Different forms of external potentials, such as constant
forces and periodically modulated forces, are now supported as well.

\section{Implementation of the communication algorithm}
\label{sec:comm}
References \cite{Anderson2008} and \cite{Anderson2013a} describe the general
data layout and GPU kernel implementation. We focus on a description of the
communication capabilities in this work.

\subsection{General strategy}
\label{sec:comm_strategy}
We implement domain decomposition in HOOMD-blue within the existing
CUDA/C++/Python framework.  The single-GPU version of HOOMD-blue does not
transfer any significant data between the host and the device during
simulation.  In the multi-GPU version, data transfer occurs at instants of
communication on every time step. Communication is an optional feature in
HOOMD-blue, and does not compromise the performance of the highly optimized
single-GPU code path. HOOMD-blue runs the single-GPU code path when executed on
one MPI rank, which results in apparently low scaling efficiency when executing
on two MPI ranks, since this additionally activates the communication routines.
However, the communication routines are designed to perform well on many, up to
several thousands, of GPUs. Communication occurs only along the necessary dimensions,
i.e. the dimension along which the domain is split into sub-domains, and the
usual minimum image convention for periodic boundaries is employed along the
other dimensions.  The domains are decomposed according to a minimum
interface-area rule.

HOOMD-blue can optionally take advantage of CUDA-aware MPI libraries (by
enabling the compilation setting \mbox{ENABLE\_MPI\_CUDA=ON}).  At the time of
writing, this includes MVAPICH2 (version 1.9 and later), and OpenMPI (version
1.7 and later). We exploit these features to optimize the communication of
ghost particles because that is the largest communication bottleneck. Given a
GPU device memory address, the MPI library can then implement any hardware- (or
software) based optimizations to accelerate data transfer between the GPU and a
second GPU or the network interface card (NIC), including RDMA-accelerated
peer-to-peer copying and GPUDirect RDMA \cite{nvidiacudaawarempi}.

\subsection{Domain decomposition}
\label{sec:domain_decomposition}
\begin{figure}
\includegraphics[width=\columnwidth]{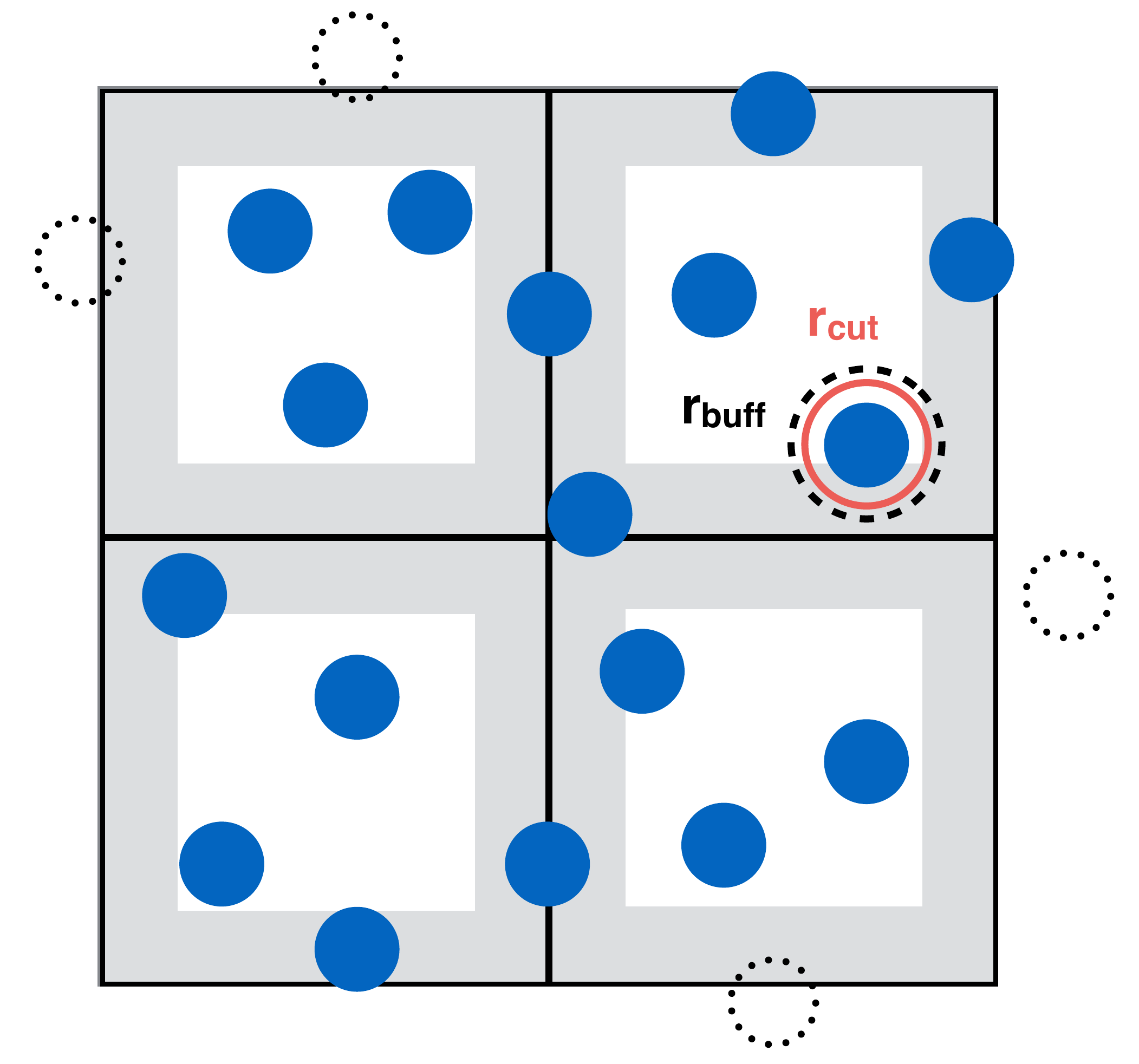}
\caption{(color online) Domain decomposition with ghost particles and periodic boundary
conditions. Shaded regions at the domain boundaries correspond to ghost
particle layers.  Short-ranged pair forces require a neighbor cut-off distance
$r_{\mathrm{cut}}$ and a buffer length $r_{\mathrm{buff}}$, and their sum
defines the ghost layer width.}
\label{fig:domain_decomposition}
\end{figure}

Domain decomposition and inter-node communication in HOOMD-blue 1.0 follow the
same basic approach as in the LAMMPS \cite{Plimpton1995} molecular dynamics
simulator: a spatial domain decomposition on a regular one-, two or
three-dimensional processor grid is employed, and boundary (or so called
'ghost') particles are communicated to compute the short-ranged pair
forces\footnote{The current version of HOOMD-blue supports only short-ranged
molecular dynamics force fields in multi-GPU mode.}
(Fig.~\ref{fig:domain_decomposition}). Particle migration transfers ownership of
particles between neighboring domains and occurs with every rebuild of the
neighbor list. Neighbor list builds occur when particles move more than
half of the buffer length to guarantee correct force computations.
The sum of cut-off and buffer length sets the width of the ghost layer, which
becomes invalid when the neighbor list is rebuilt.  Between neighbor list
builds, particles may move beyond the boundaries of the local domain without
migration. Positions (and, depending on the force field,
velocities or orientations) of ghost particles are updated every time
step, which requires communication.

\subsection{GPU-based communication algorithm}
\label{sec:gpu_communication_algorithm}
The core of HOOMD-blue's new MPI capabilities is the communication algorithm,
which features a number of optimizations to reduce unnecessary device-host data
transfers, and in particular allows sending/receiving particle data directly
to/from the device-side data structures. The basic particle packing and
unpacking routines are fully implemented on the GPU.

\subsubsection{Particle migration}
\label{sec:particle_migration}
Particle data is stored in a structure of dense arrays for optimal execution of
communication kernels on the GPU. When particles are migrated, the particle data
is reordered by first removing particles that leave the domain, which creates
holes in the arrays. These are removed using a stream compaction based on the
scan primitive \cite{ModernGPU}.  Subsequently new particles are added at the
end of the arrays. Even though the simultaneous reordering of the GPU arrays
for position, velocity, acceleration etc.\  is a relatively expensive operation,
it is performed only infrequently, typically every 5-20 time steps
when the neighbor list needs to be rebuilt.

In LAMMPS, the exchange with the neighboring domains is reduced to six
communication calls, including exchanges along the east-west, north-south, and
up-down directions, which is the minimum required in three spatial dimensions
\cite{Plimpton1995}. The approach bundles several smaller messages into fewer
large messages, to increase communication bandwidth utilization. For GPU
implementation, however, this is a disadvantage, since it introduces
data-dependency between the exchanges, as the receive buffers for particles
from, say, an eastern neighbor need to be scanned for particles that must be
forwarded in the next exchange in the north direction. In particular, an
extra device-host copy and a kernel call between every exchange is incurred.
which increases latency. We determine that simultaneously
communicating with all 26 neighbors using non-blocking MPI calls performs better
in all cases. The MPI library can optimally pipeline between those
transfers.

To limit data transferred over PCIe, we pack the buffer of particles to send on
the GPU in a device-gather operation. The data, which represents only a small
subset of particles in a thin slab near the boundary of every domain, is copied
to the host and the host buffer is then passed to the MPI library. Receiving
particles works analogously. We generally use mapped and pinned memory for host
buffers to obtain higher bandwidth and compute-data transfer overlap. In some
circumstances, however, non-mapped (explicitly staged) memory is faster. We
evaluate the performance of both modes at run time and select the fastest using
auto-tuning (see Sec.~\ref{sec:autotuner}).

\subsubsection{Ghost particle exchange}
\label{sec:ghost-nonbonded}
Ghost particles are exchanged after particle migration. They are appended to
the particle data arrays and remain valid until the next exchange. A ghost
particle can be simultaneously a neighbor in up to three domains - one 'face'
neighbor, one 'edge' neighbor and one 'corner' neighbor. The same particle
needs to be copied into up to three output buffers.  While this represents a
straight-forward operation to do on the CPU, efficient parallel replication of
ghost particles on the GPU requires a more elaborate implementation.  We
implement particle replication as summarized in Fig.~\ref{fig:ghosts}.  First,
a set of flags (a bitfield) is initialized for every particle that indicate
which neighboring directions the ghost particle is supposed to be sent to (such
as: north-west).  Second, the algorithm counts the number of neighboring boxes the
particle is replicated to and writes pairs of particle index and destination
rank for every replication into an intermediate buffer. Multiple destination
ranks can map to the same particle index, so multiple output elements are
generated per input element.  To implement such an operation efficiently using
an atomics-free algorithm, i.e.\ an algorithm which does not rely on
random ordering of parallel read/write accesses to the same location,
we turn to the load-balancing search or 'expand'
algorithm within the ModernGPU framework \cite{ModernGPU}. We adapt this
routine to perform both the expansion of particle indices and the determination
of neighbor ranks in a single kernel call.  Subsequently, output elements are
sorted by destination rank (using the MergeSort algorithm \cite{ModernGPU}),
and the begin and end indices for each destination rank are found using
vectorized binary search and mark the boundaries of the send buffer. These
optimizations resulted in significant speed-up of the ghost exchange over an
implementation that involves many different kernel calls (e.g.  using Thrust
\cite{Bell2012}).

\begin{figure}
\begin{framed}
{\bf Ghost particle exchange:}
\begin{enumerate}
\item Mark ghost atoms for sending
\item {\em In one kernel}, determine destination ranks for every ghost particle
and write out (particle index, destination rank) pairs into a ghost particle
list where a particle index can occur multiple times, using the load-balancing
'expand' operation
\item sort list by destination rank and determine start and end indices of
every rank
\item gather particle positions (velocities, ..) of ghost atoms according to
ghost particle list into send buffer
\end{enumerate}
\end{framed}
\caption{Ghost particle exchange optimized using load-balancing search
\cite{ModernGPU}.}
\label{fig:ghosts}
\end{figure}

\subsubsection{Overlapping communication and computation and other optimizations}
\label{sec:overlap}
We evaluated two possible strategies for overlapping communication and
computation, (i) splitting the force computation into particles with local- and
nonlocal neighbors, and (ii) overlapping collective MPI calls with computation.

In the first approach, the force computation for every local particle is split
into the forces due to local particles, and forces due to ghost particles. The
computation of forces due to interactions with local particles can be
overlapped with the communication of ghost particles, after the completion of
which the forces due to ghost particles are computed in a separate kernel
call.  However, because this approach requires two (instead of one) invocations
of the force kernel, it incurs launch overhead for the extra kernel. We tested
this approach and find that is more expensive than a non-overlapping,
single-kernel approach.

In the second approach, forces are always computed using the current particle
neighbors, before the results of the neighbor list distance check are known.
Since the distance check requires a global synchronization (an MPI collective),
this synchronization can now be performed simultaneously with the force
computation. When the distance check returns positive, the precomputed force
must be discarded, particles migrated, and then correct forces computed.  For
typical benchmarks where particle migration occurs every 5-20 time steps, the
proposed scheme results in better performance.  Because theoretially,
situations can occur in which performance is adversely affected, in particular
for short neighbor list rebuild intervals, the performance difference is
measured at run-time and compute/communication overlap is auto-tuned.

A further optimization of the communication pattern is to group global
synchronization calls and neighborhood communication into different
phases of the time step. Examples for global synchronization include
broadcast of parameters during NVT integration, or MPI reduction calls for
computation of thermodynamic properties such as kinetic energy, virial etc.
Call-back slots are provided in the communication algorithm for different HOOMD
classes to opt-in to the optimized execution pattern, to the extent this is
possible given data-dependencies. The grouping improves performance because it
mitigates the effect of load imbalances, and global synchronization occurs only
at one instant during the time step, instead of being interleaved with
computation or neighborhood communication.

Because our communication algorithm does not make use of atomic operations, it
is fully deterministic, i.e.\ the simulation results are bit-wise reproducible
(e.g. for debugging). However, HOOMD-blue's cell list algorithm uses atomic
operations, and overall execution is therefore non-deterministic. We stress that
non-determinism arises here as a side-effect of parallel execution (that can
sometimes be avoided), due to the non-associativity of floating point
operations at finite numerical precision. Since MD integrators are affected by
numerical precision in the same way, lack of determinism does not imply
degraded stability in a thermodynamic sense.

As a minor optimization, we map the Cartesian rank index to a logical MPI rank,
so as to group spatial domains of the decomposition
together that are executed on the same physical node.  For the neighborhood
exchanges this reduces inter-node communication, which is more expensive
than intra-node communication.

\subsubsection{Bonded forces}
In HOOMD-blue 1.0, two-particle bonds, three-particle angles and four-particle
dihedrals and impropers are treated as instantiations of a generic C++ class
template, {\tt BondedGroup}, to avoid code duplication for different types of
particle groups with small numbers of members. For simplicity, we refer to a
'bonded group' as a bond in what follows, implying that the description
applies to angles, dihedrals and impropers in an analogous way. A bond
is simply an unordered pair of particle IDs, together with a bond type number.

Bond data structures are dynamic, and bonds can be added or removed during the
simulation using the python scripting interface. Also, in MPI simulations,
bonds are stored {\em locally} and are migrated between domains together
with the particles, which ensures $\mathcal{O}(N/P)$ memory and compute scaling
for bonds. We developed an efficient communication algorithm for bonded particles.
Remarkably, our algorithm does not explicitly require specifying a
maximum bond-length, which would otherwise be used to simplify the treatment of
ghost particles, because bond neighbors would be treated in the same way as
non-bonded neighbors by simply widening the ghost layer to the maximum bond
length. For long or very floppy bonds, this would result in the communication
of too many non-bonded neighbors.

\begin{figure}
\begin{framed}
{\bf Bond migration:}
\begin{enumerate}
\item ({\em compute})
Mark particles that leave the local domain for sending (also used for particle migration)
\item ({\em communicate})
For all particles leaving the domain that are members of incomplete bonds,
update neighbors with destination processor rank \label{update_rank}
\item ({\em communicate})
For all particles leaving the domain and members of bonds, copy bonds to destination rank
\item ({\em compute})
Prune bonds that have no local members
\end{enumerate}
{\bf Bonded ghost particle exchange:}
\begin{enumerate}
\setcounter{enumi}{5}
\item ({\em communicate}) Using the table constructed or updated in step \ref{update_rank},
exchange ghost particles with neighbors
\end{enumerate}
\end{framed}
\caption{Communication of bonded groups (bonds, angles and dihedrals).}
\label{fig:bonds}
\end{figure}

In our algorithm (summarized in Fig.~\ref{fig:bonds}), bond migration occurs
directly before particle migration, but after local particles have been identified
that leave the domain. The difference between the two steps, besides
the difference in the data objects that are communicated, is that particles are
uniquely local to a single domain but bonds can span two domains. If a bond
is split because one of its members migrates to a neighboring domain,
the bond is stored simultaneously in both processors' data structures, and a
bond is only removed from a processor's storage if none of its particle members
are local anymore. At the time the system is initialized, bonds are distributed
onto processors.

The issue of correct force computation is a little more involved, because it is
necessary to exchange ghost particles with a neighboring domain if they share a
split bond.  However, the sending processor does not know {\em a-priori} where
to send the ghost particles, since the bond can span any of the processor
boundaries. In particular, the fact that a bond member is in the proximity of a
boundary does not imply that the bond actually spans that boundary. To account
for this issue, it is necessary to keep track of the processor ranks that share
the bond, using a data structure to store all processor ID's that own any
member particles. This data structure is initialized at the beginning of the
simulation using a broadcast operation between neighbors. It is then updated
with every particle  migration, by exchanging information with the processors
that own other bond members. Hence, bond migration consists of two
communication steps, where in the first step we propagate the IDs of particles
that leave their domain to the neighboring processors, allowing them to update
their rank tables, and in the second step we actually migrate (or replicate)
the bonds.

Using current information about bond member ownership, it is straightforward to
identify the neighbors with which ghost particles have to be exchanged in order
to compute the bond force. Particles are marked for sending in a specific
direction, e.g. up-north-west, based on that information. Subsequently, these
ghost particles are treated the same way as nonbonded ghost particles, see Sec.
\ref{sec:ghost-nonbonded}.

Communication only occurs between neighboring domains, so
bonds cannot span more than two domains, and because of the periodic boundary
conditions bond lengths are limited to half the local domain size.

\section{Optimizations for strong scaling}
\label{sec:opt_strong_scaling}
Ideal $N/P$ scaling of a multi-processor simulation of $N$ particles on $P$
GPUs not only requires low-latency communication routines, but also that all parts
of the computation exhibit, linear $N/P$ scaling. As the workload $N/P$ on a
single GPU decreases with increasing numbers of processors at constant total
system size $N$ in the strong scaling limit, optimal kernel launch parameters
for the pair potential kernels change. These low-level parameters define
how the {\em Single-Instruction-Multiple-Threads} (SIMT) model maps onto the GPU
workload. In previous versions of HOOMD-blue one particle was mapped onto one
GPU thread, which ceases to be efficient at low numbers of particles ($N/P
\lesssim 20,000$) because the GPU is underutilized. Therefore, we now assign
multiple threads to a particle \cite{Brown2012}.

\subsection{Force computation with a cooperative thread array}
\label{sec:opt_strong_scaling_force}
\label{sec:nlist}
\begin{figure}
\begin{framed}
{\bf for all particle $i$ in groups of $w$ parallel threads do:}
\begin{enumerate}
\item Look up cell neighbors of particle $i$
\item Iterate over cell neighbors $k$ with a stride of $w$ (block thread $j < w$ processes
neighbor with offset $j$)
\item Set $\mathtt{is\_neighbor}=1$ if particle is a neighbor ($r < r_{\mathrm{cut}}$)
\item {\tt o} = $j$th element of {\em w}-wide exclusive prefix sum over {\tt has\_neighbor}
\item {\tt n} = result of reduction ($j+1$th element of prefix sum)
\item If $\mathtt{is\_neighbor}> 0$, write neighbor index $k$ into neighbor list at position $\mathtt{n\_tot + n}$
\item {\tt n\_tot += n}
\item Store {\tt n\_tot}
\end{enumerate}
\end{framed}
\caption{Neighbor list algorithm using $w$ cooperative threads
per particle.}
\label{fig:nlist}
\end{figure}

Instead of processing all particle neighbors in a single loop per thread, we
iterate over nonbonded particle neighbors using an array of $w$ cooperative
threads. The parameter $w$ must be less than or equal to 32 in the current GPU
generation because this is the number of threads that are executed in a warp
and can communicate with the warp shuffle instruction.  In HOOMD-blue 1.0, both
the neighbor list and the force computation kernels therefore take $w$ as an
additional input parameter, which controls the size of the block.  Here,
$w=2^m$ is a power of two, and if $m < 5$ a single warp is split into multiple
segments.

The pseudo-code for the neighbor list algorithm is shown in
Fig.~\ref{fig:nlist}.  The algorithm runs in parallel in blocks of width $w$,
and loops through the cell neighbors of every particle in a strided fashion
(with offset equal to the offset of the thread in the block), to compute the
neighbors that are within the cut-off distance. The final result is obtained by
compacting all neighbor particle indices into one (output) row of the neighbor
list, using intra-warp stream compaction. Threads of a warp
segment can communicate with each other via the fast {\tt \_\_shfl} intrinsic
(on Kepler) or via shared memory (on Fermi).

Similarly, for the force computation the work for particle $i$ is parallelized
over its neighbors, processing $w$ elements of its neighbor list in parallel,
again in a strided fashion. Each of these threads contributes to the total
force on particle $i$. After the partial forces have been computed by the $w$
threads, an intra-warp floating point reduction is applied
to total up the particle force, which then is written to global memory.

\subsection{Autotuner}
\label{sec:autotuner}
Both the neighbor list and the force computation kernels now depend on $w$, in
addition to the thread block size, which is another parameter in the SIMT
execution model equal to the number of threads that share the same thread-local
memory. However, the block size is only relevant for performance.  Because we
now have two parameters to tune, it is no longer practical to manually select
the optimal parameters  out of $160=32 \times 5$ (block size $b= n\times32$
$b \le 1024$ \cite{nvidiacudaprogrammingguide}, and $w = 2^m$ with $m \le 5$)
possible combinations. Moreover, the optimal parameters may change in
a single simulation, as a function of density or the pair potential parameters.

Our solution is to implement an autotuner, which measures kernel execution
time for every parameter. The tuner sweeps through the possible parameter
combinations {\em during the simulation}, and compares the median run time from
five different kernel executions with the same parameters among different
parameter sets.  After having found the value that results in the shortest
execution time, this optimal value is fixed for a predefine number of MD time
steps.  The tuning typically takes between 15,000 and 20,000 steps to complete,
which in most cases represents a small fraction of the overall number of steps.
Periodically, the tuner scans through the parameters to re-adjust the optimal
parameter setting.

\subsection{Scaling of compute kernels vs. communication}

\begin{figure}
\centering\includegraphics[width=\columnwidth]{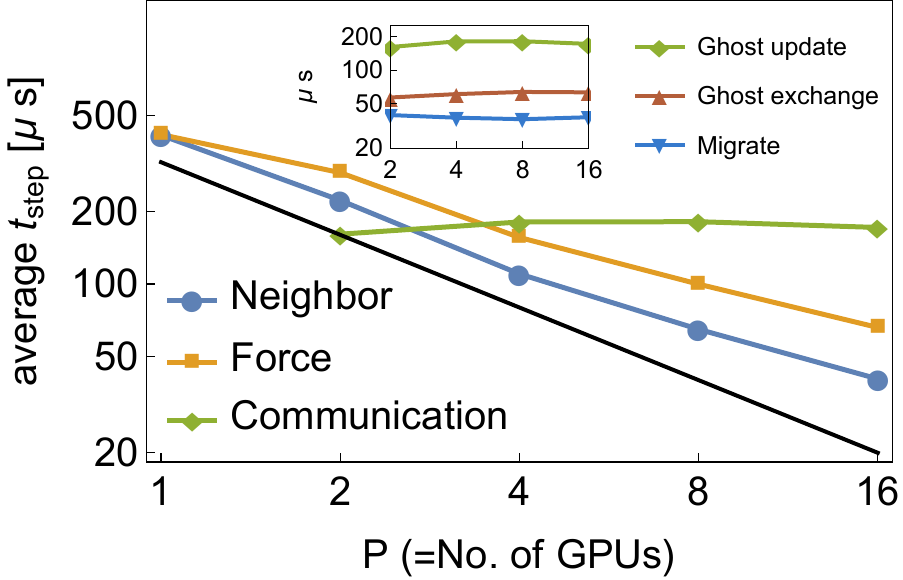}
%\centering\includegraphics[width=\columnwidth]{comm_scaling}
\caption{(color online) Time $t_{\mathrm{step}} = t_{\mathrm{loop}}/n_{\mathrm{steps}}$ in
individual parts of the algorithm for a $N=64,000$ Lennard-Jones liquid
benchmark on $P=1\dots 16$ K20X GPUs, showing contribution of pair force,
neighbor list and dominant communication (ghost update) routines.
{\em Inset:} Contribution of the migration, ghost exchange and update phases
to overall time spent in communication.}
\label{fig:kernel_scaling}
\end{figure}

We analyze the performance of individual parts of the simulation in
Fig.~\ref{fig:kernel_scaling} (upper panel).  The main plot shows the time
spent in different subroutines in a multi-GPU simulation as a function of the
number of GPUs. As is evident from that plot, the computationally most
expensive subroutines -- the neighbor list construction and the pair force
kernels -- scale essentially inversely linear with the number of GPUs,
$t_{\mathrm{step}} \sim N/P$ ($N=64,000=const.$), down to (at least) $4,000$
particles per GPU, as a result of the optimizations for strong scaling
described in Sec.~\ref{sec:opt_strong_scaling_force} and auto-tuning.  However,
the time spent in communication (green curve), which occurs every time step, is
almost independent of the number of particles per GPU, imposing constant
latency every simulation time step.  Particularly, we infer from the cross-over
of the two curves between $P=4\dots8$, that scaling {\em must} break down at
fewer than $\sim 10,000$ particles per GPU, where communication becomes the
limiting step.

We further detail the contribution of the three phases of communication (see
Sec.~\ref{sec:gpu_communication_algorithm}) to the average time per step in the
inset of Fig.~\ref{fig:kernel_scaling}, and while all contributions depend only
weakly (if at all) on the number of particles per GPU in the regime studied
here, the dominant amount of time is spent inside the ghost update occurring
every time step. It is therefore this part of the communication that we heavily
optimize, through overlap with global synchronization calls
(Sec.~\ref{sec:overlap}) and exploitation of CUDA-aware MPI calls
(Sec.~\ref{sec:comm_strategy} and below).

\section{Scaling performance}
\label{sec:scaling_titan}

\subsection{Weak scaling}
\label{sec:weak_scaling}
A first requirement for a scaling code is its capability of handling large
simulation workloads, which requires $\mathcal{O}(N/P)$ memory scaling.  We
perform a weak-scaling benchmark of a LJ fluid on the ORNL Titan Cray XK7, keeping
the number of particles per GPU constant at $N/P=32,000$.  The reasons for the
choice of the rather simple LJ benchmark, instead of benchmarking e.g.  a
molecular system, are that it is (i) computationally inexpensive, and (ii) a
standard benchmark for MD.  Because of (i), any performance issues with the
communication algorithm will be exposed, and because of (ii) we can compare,
e.g., to the LAMMPS software package with its similar feature set. We perform
the benchmark in the constant temperature (NVT) ensemble with a Nos\'e-Hoover
thermostat, albeit one that is not completely symplectic\footnote{A symplectic
MTK integrator for NVT was introduced in HOOMD-blue 1.0.1 for improved
stability over the the original Nos\'e-Hoover NVT thermostat, which is
deprecated in that version. The symplectic integrator requires multiple
collective synchronization calls per time step.}, and which performs one
collective synchronization for kinetic energy summation every time step.

In particular, we simulate a fluid of up to $N=108,000,000$ particles on 3,375
GPUs.  The system has been equilibrated from a random initial configuration
over $100,000$ time steps, at packing fraction $0.2$, and with a time step of
$\delta t = 0.002$ (all quantities in self-consistent or LJ units). The
particles interact via the 12-6 LJ potential with $\sigma=1.0$, cut off at
$3.0\sigma$.  The interaction parameter is $\epsilon=1.0$.  The buffer length
$r_{\mathrm{buff}}$ is determined as the value that yields the highest number
of time steps per second in a preliminary tuning run, and the distance check
criterion is applied with the minimum frequency at which no dangerous builds
are observed. Subsequent to an auto-tuner warm-up period of 20,000 time steps,
the simulation performance we report is averaged over 10,000 steps.

The corresponding LAMMPS-GPU benchmark was performed on the same machine,
starting from the same equilibrated initial configuration, as well as the same
simulation parameters, but the optimal configuration was determined to have a
fixed buffer size (skin length) of $0.4$, and neighbor lists were rebuilt every
5 time steps. LAMMPS-GPU performance benefits from using more than one CPU core
to drive the single GPU per node. Therefore, the number of MPI ranks per node
was varied between 1 and 16.  For every number of nodes, the fastest value of
the number of processes per node was selected.  To facilitate the
initialization of such large systems, the cubic simulation box configuration
was replicated along three spatial dimensions, using LAMMPS' {\em replicate}
command, or HOOMD-blue's {\em system.replicate()} method.

\begin{figure}
\includegraphics[width=\columnwidth]{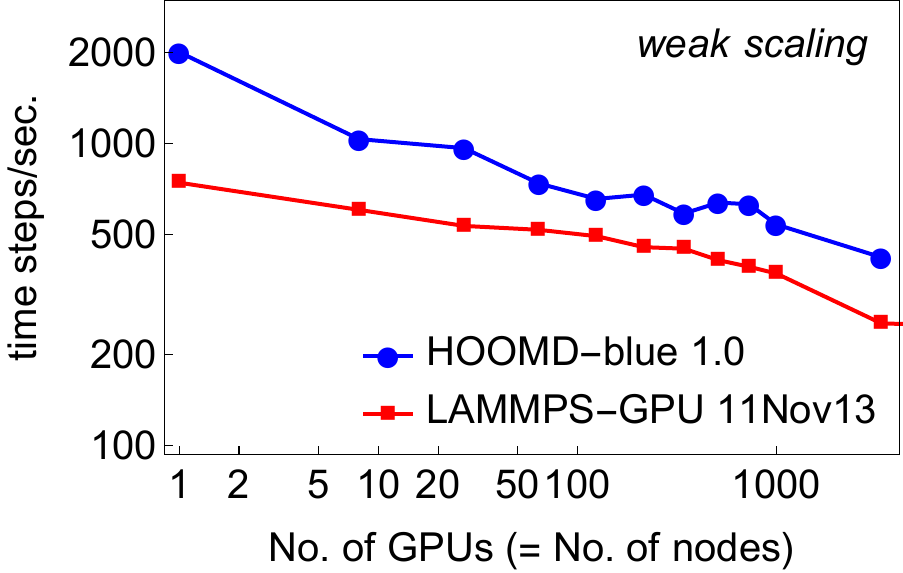}
\caption{(color online) Weak scaling performance of HOOMD-blue (solid circles) and LAMMPS-GPU
(solid squares) on Titan, in number of time steps per second vs. number of
GPUs, for a constant number of particles per GPU $N/P=32,000$, for
$1\dots3,375$ GPUs.}
\label{fig:weak_scaling}
\end{figure}

Figure \ref{fig:weak_scaling} shows the number of time steps per second as a
function of the number of GPUs used in the weak scaling simulation, for
HOOMD-blue and LAMMPS-GPU. Data is shown for up to the maximum total system
size we succeeded to initialize on Titan using HOOMD-blue. Ideal weak scaling
corresponds to constant performance as a function of the number of GPUs. As
expected, the largest drop in performance for HOOMD-blue occurs when the number
of GPUs increases from one to $\sim 8$ GPUs, which corresponds to switching
from no communication to a fully three-dimensional ($2\times 2\times 2$) domain
decomposition. The further decrease in performance of only about 50\% over a
range of over 3,000 GPUs is likely caused by increased communication latency
arising from collective MPI calls. HOOMD-blue and LAMMPS-GPU scaling are in
otherwise good qualitative agreement, indicating the absence of major
bottlenecks in either code.  HOOMD-blue performance is superior to that of
LAMMPS-GPU by roughly a factor of two, which impressively demonstrates the
speed-up gained by optimizing for the GPU exclusively and highlights some
present limitations of a hybrid GPU/CPU approach. The hybrid approach used by
LAMMPS relies on offloading only the compute-intensive force computation to the
GPU and on using the GPU in a sparse fashion. In turn, for maximum performance
GPU kernel launches from several MPI ranks per node are multiplexed onto the
same GPU (using the Hyper-Q capability of NVIDIA Kepler GPUs), which can
create additional kernel launch overhead. For HOOMD-blue, we varied the number
of CPU cores assigned to a GPU and did not find any benefit from using more
than one core per GPU.

\subsection{Strong scaling}
\label{sec:strong_scaling}
We measured HOOMD-blue performance for the strong scaling of a simple LJ liquid
benchmark of $N=10,976,000$ particles (for the benchmark script, see
\cite{supplementary}) on Titan.  We compared HOOMD-blue performance to the
performance of the LAMMPS-GPU package, and the details of the simulation are
otherwise the same as for the weak scaling.

\begin{figure}
\includegraphics[width=\columnwidth]{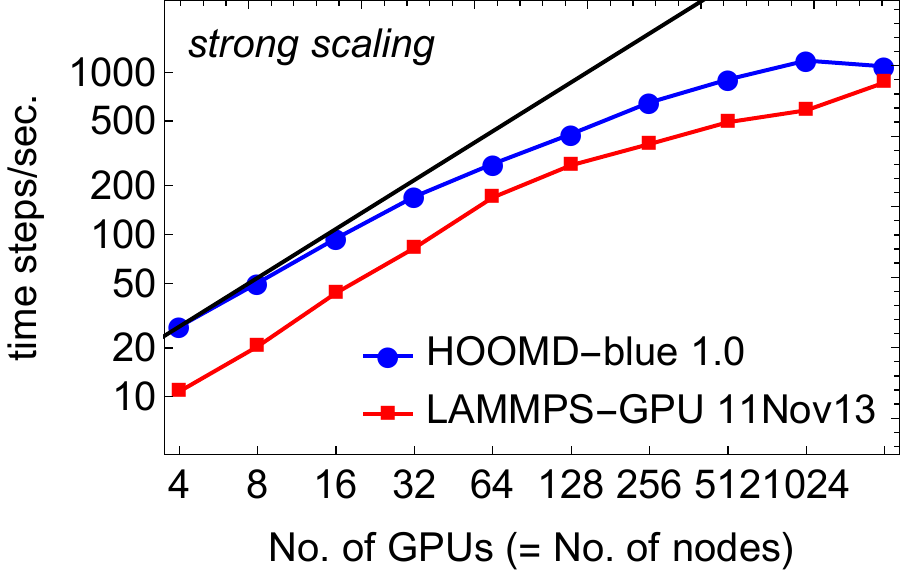}
\caption{(color online) Strong scaling of a LJ liquid benchmark with $N=10,976,000$ particles
(single precision) on the Titan supercomputer (Cray XK7, Oak Ridge National
Laboratories). Shown is the performance in terms of number of time steps per
second (TPS) for HOOMD-blue (circles) and LAMMPS-GPU (squares), running on
$P=4\dots 2048$ GPUs. HOOMD-blue performance is optimal with 1 MPI rank per
node/GPU, for LAMMPS-GPU up to 16 CPU cores are assigned to one device. The
solid line shows ideal linear scaling.}
\label{fig:titan_strong}
\end{figure}

Figure \ref{fig:titan_strong} shows HOOMD-blue performance, i.e.\ number of time
steps per second vs. number of GPUs, compared to LAMMPS-GPU for the LJ liquid
benchmark, on up to 2,048 GPUs. As can be seen, essentially ideal scaling holds
up to 32 GPUs, beyond which the scaling shows decreased efficiency. However,
scaling holds up to 1,024 GPUs in this benchmark. Both HOOMD-blue and
LAMMPS-GPU scale qualitatively similarly, such that they are limited by
architecture-inherent communication bottlenecks in the same way. However, as
for weak scaling (Sec.~\ref{sec:weak_scaling}), HOOMD-blue achieves a roughly constant,
two-fold speed-up over the LAMMPS-GPU code, demonstrating the
effectiveness of the above optimizations.

%\begin{figure}
%\centering\includegraphics[width=\columnwidth]{timestep}
%\caption{Profile of a time step}
%\end{figure}

\subsection{Scaling efficiency}
\label{sec:efficiency}
\begin{figure}
\includegraphics[width=\columnwidth]{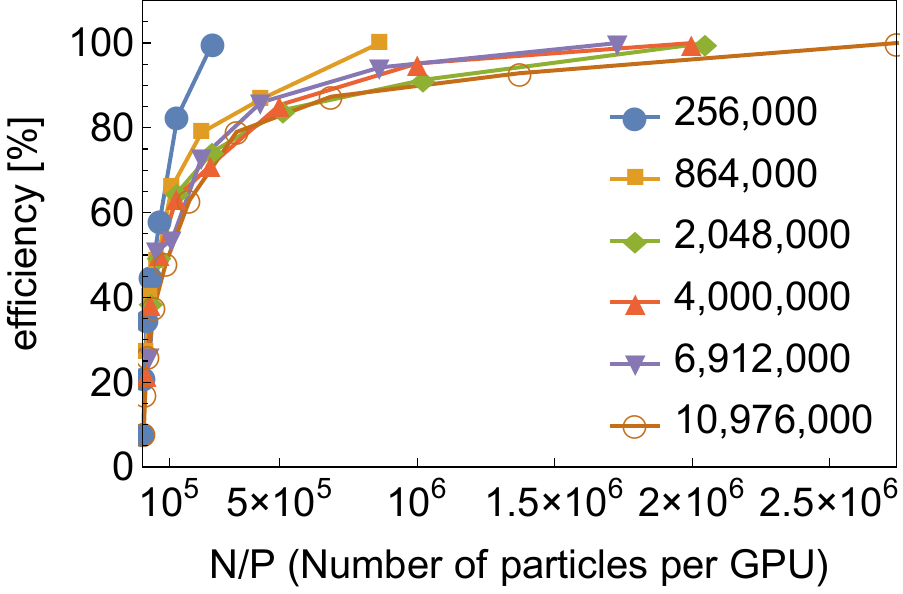}
\caption{(color online) Strong scaling efficiency of HOOMD-blue in a LJ liquid benchmark for
seven different total system sizes between 256,000 and 10,976,000 particles
(single precision), running on at least $P_{\mathrm{min}}$ up to 2048 GPUs (for
the largest system size) of Titan.  Shown is the normalized performance
$\mathrm{TPS}(P)\, P_{\mathrm{min}}/P\, \mathrm{TPS}(P=P_{\mathrm{min}})$ vs.
number of particles per GPU $N/P$.}

\label{fig:efficiency}
\end{figure}

Only in rare circumstances, however, does the maximum achievable performance
matter; more often the compute resources are limited and shared with other
researchers.  It is therefore relevant to know what is the maximum number of
GPUs a simulation can be run with to reach a prescribed minimum scaling
efficiency. To this end, we compared strong scaling efficiency at various total
system sizes, for the same system as studied in Sec.~\ref{sec:strong_scaling}.
The result is shown in Fig.~\ref{fig:efficiency} for six different total system
sizes between $N=256,000$ and $N=10,976,000$ particles, on different numbers of
GPUs up to 2,048 GPUs.  It turns out that the strong scaling performance is
essentially independent of the total size. With the slight exception of the
$P=1$ and $P=2$ data points for the smallest system ($N=256,000$), for which we
expect the largest relative amount of communication overheads, the data falls
onto a master curve obtained from plotting all results as scaling efficiency
$\gamma = \mathrm{TPS}(P)/P\, \mathrm{TPS}(P=1)$ vs. number of particles $N/P$
per GPU. In cases where at least $P_{\mathrm{min}}>1$ GPUs are required
to hold the system in memory, we normalize by $\mathrm{TPS}(P_{\mathrm{min}})/P_{\mathrm{min}}$.

The data in Fig.~\ref{fig:efficiency} shows that maximum scaling efficiency
($\gamma \sim 1$) is achieved for $N/P \gtrsim 10^6$. However, a very good
efficiency of $70\%$ is already achieved with $N/P \sim 200,000$ particles
per GPU, which is easily reached by many medium- to large scale simulations
e.g. of polymeric liquids.

\section{Polymer brush scaling benchmark}
\label{sec:tethers}
\begin{figure}
\centering\includegraphics[width=\columnwidth]{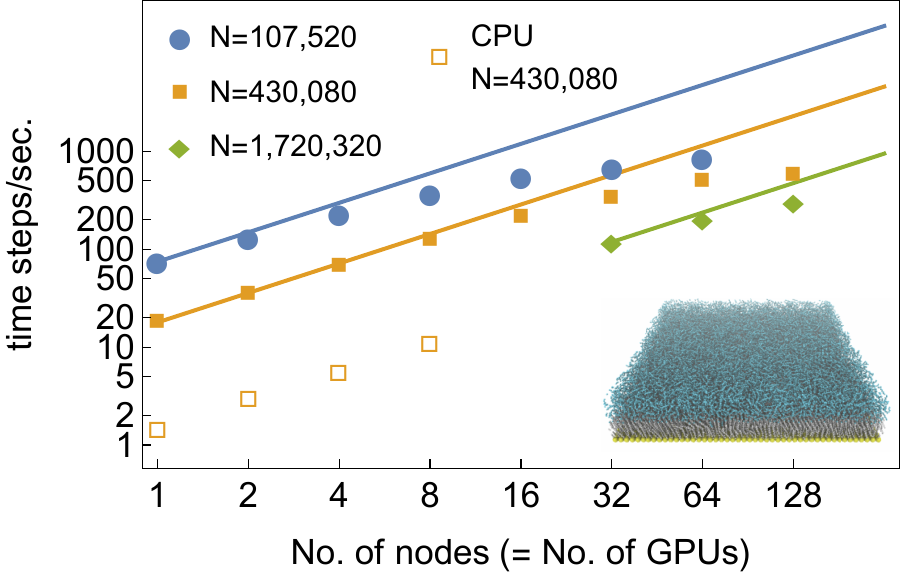}
\caption{(color online) Benchmark for an oleic acid brush in hexane solvent.
Shown is the performance in terms of number of time steps per second vs.
number of GPUs or number of CPU sockets for three different total system sizes.
Solid symbols: GPU benchmark at different total system sizes, open symbols: CPU
benchmark at N=430,080.  Because of memory limitations, the smallest number of
GPUs for the system of $1,720,320$ particles is 32, and because of the minimum
size required for the subdomain in a spatial domain decomposition, the largest
number of nodes in the CPU benchmark is eight. Solid lines show ideal
linear scaling. }
\label{fig:tethers_scaling}
\end{figure}

We also analyze the scaling performance of a more complex benchmark system, a
polymer brush in presence of a polymeric solvent. Chemical details
are described at the level of a united atom model \cite{Paul1995}. The solvent
is hexane, and the tethers are oleic acid molecules, grafted onto an
immobile wall. The system is used to study effective interactions of
polymer-tethered surfaces \cite{Millan:unpub}.  Here we use it as a model
system to study HOOMD-blue performance under conditions that are more
computationally demanding than a LJ liquid.

The polymer backbones are parameterized using bond, angle and dihedral
potentials, and the system is equilibrated at constant pressure and temperature
(NPT ensemble). For this bennchmark, we replicate the initial configuration
along the wall (or x-y) dimension using {\em system.replicate()}, to produce
three different total system sizes (numbers of polymeric particles) of
$N=$107,520, 430,080 and 1,720,320.  Detailed parameters of the benchmark,
which was run on Blue Waters, are documented in the supplementary information
\cite{supplementary}.  As shown in Fig.~\ref{fig:tethers_scaling}, the
performance (time steps / second) scales approximately linearly with the number
of GPUs, up to a maximum number of GPUs that increases with total system size.
For example, for the system of 430,080 polymer beads, saturation sets in at 32
GPUs, corresponding to 13,440 particles per GPU. For this system, very good
scaling is observed and a maximum speed-up of a factor $\approx 32$ over
single-GPU execution is achieved,  which shows that the speed-ups obtained in
the LJ benchmarks are robust and transferable to more complex interaction
models.

Moreover, we compare the GPU to CPU performance of HOOMD-blue running on the 16 CPU cores
of Blue Waters' XK7 nodes (1 AMD 6276 Interlagos CPU, 16 integer scheduling units,
nominal clock speed 2.3 GHz), for the $N=430,080$ benchmark. The strong scaling data for time
steps per second vs. number of nodes is shown in
Fig.~\ref{fig:tethers_scaling}, open symbols, and demonstrates an average
speed-up of a factor of $\mathrm{TPS}(n\, \mathrm{GPUs})/\mathrm{TPS}(n\,
\mathrm{CPUs})\approx 12.5$.

\section{Strong scaling with GPUDirect RDMA}
\label{sec:gdr}
\subsection{Technological background}
\begin{figure}
%\begin{framed}
\centering\includegraphics[width=\columnwidth]{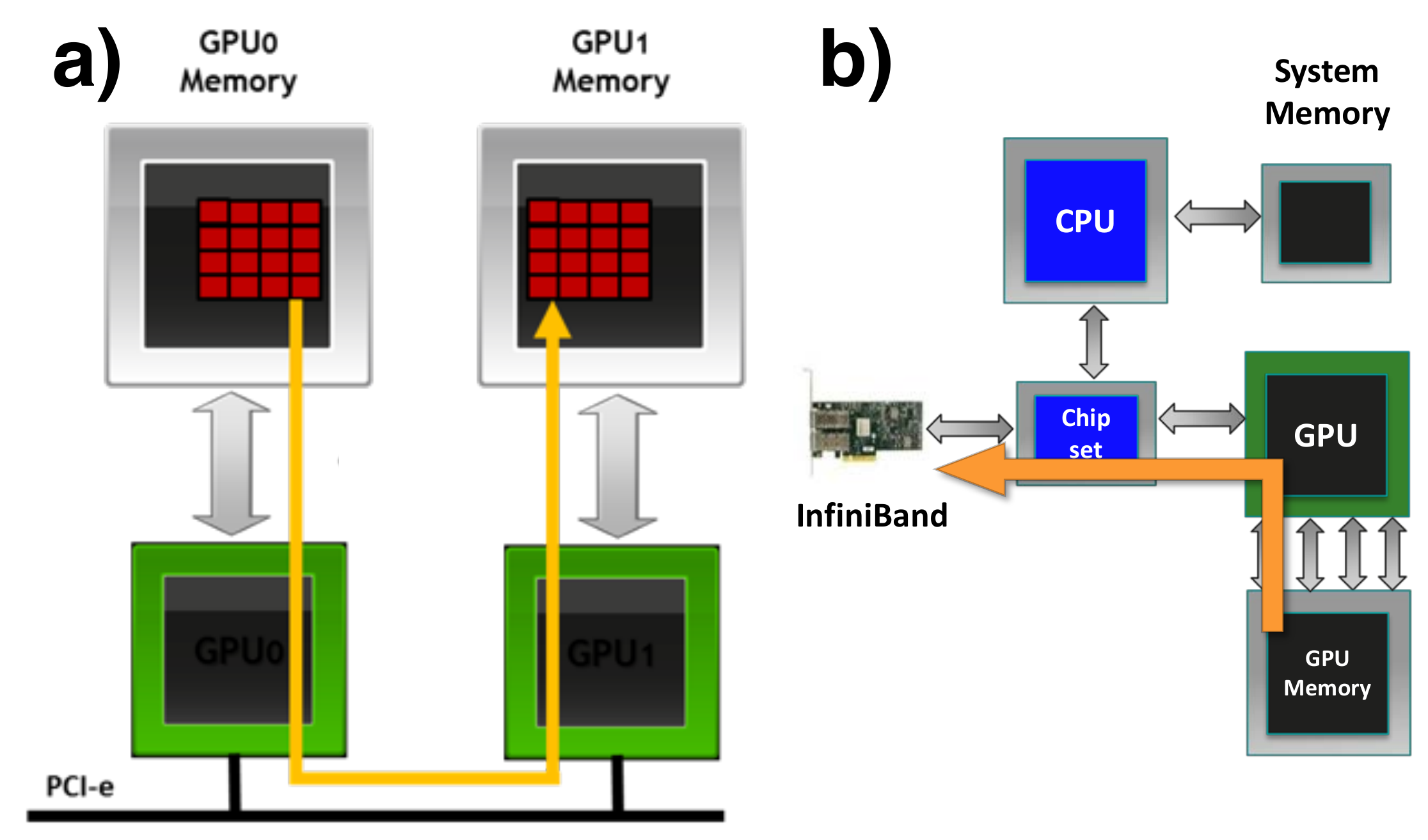}
%\end{framed}
\caption{(color online) RDMA enabled technologies for data transfer between GPUs across the
PCIe bus without involving the CPU. a) Intra-node Peer-to-peer
b) GPUDirect RDMA. Reproduced from the GPUDirect webpage \cite{nvidiagpudirect}.}
\label{fig:rdma}
\end{figure}

Since rapid developments in GPU hardware have out-paced the speeds at which
system architecture, and particularly PCIe, can transfer data (16 GB/s peak
bandwidth in PCIe generation 3), this has generated pressure on the development
of latency-optimized technologies for GPU-to-GPU communication, either over a
PCIe switch, with two GPUs connected to the same switch, or via the
intermediate of a network interface card (NIC) for Infiniband connected to the
same PCIe segment. The first generation implementation of a
hardware-accelerated communication path for GPUs was GPUDirect v1
\cite{Shainer2011}, which eliminates unnecessary buffer copies within host
memory for staging data to the NIC. To use GPUDirect, it is essentially only
required that the send/receive buffers are page-locked, which enables fast access
by the network card. However, with GPUDirect v1, the data still needs
to be available in host memory, making it necessary to copy the data to and
from there for sending and receiving.

With the release of GPUDirect v1, MPI stacks have been developed (starting with
MVAPICH2 1.8 and OpenMPI 1.7, and others) that accept memory pointers for the
staging buffers, designed to relieve the application developer from having to
manually optimize staging protocols and or pipelining techniques performance
\cite{Wang2011,Wang2014b,Potluri2013}.

The second version of the GPUDirect technology, introduced with Fermi
generation GPUs, offers the possibility of intra-node GPU-GPU transfer
using the RDMA feature of PCIe, without involving the CPU
(Fig.~\ref{fig:rdma}, left panel). Using this technique, up to $80-90\%$ of
PCIe peak bandwidth can be achieved for these transfers. However,
because  peer-to-peer transfers are limited to a single node, transfers that
involve inter-node communication e.g. over Infiband are not optimized.

The latest generation of the GPUDirect technology, GPUDirect RDMA (GDR), solves
this problem by extending RDMA access to the NIC and allows fast data transfer
between remote GPUs, completely offloading the CPU from the communication
burden (Fig.~\ref{fig:rdma}, right panel). All of these protocols are supported
in the MPI libraries MVAPICH2 2.0 GDR, and in OpenMPI 1.7.4 and later. In the
following, we analyze the efficiency and discuss limitations of this advanced
technology in the example of HOOMD-blue scaling benchmarks.

\subsection{GPUDirect RDMA benchmarks}
\begin{figure}
\centering\includegraphics[width=\columnwidth]{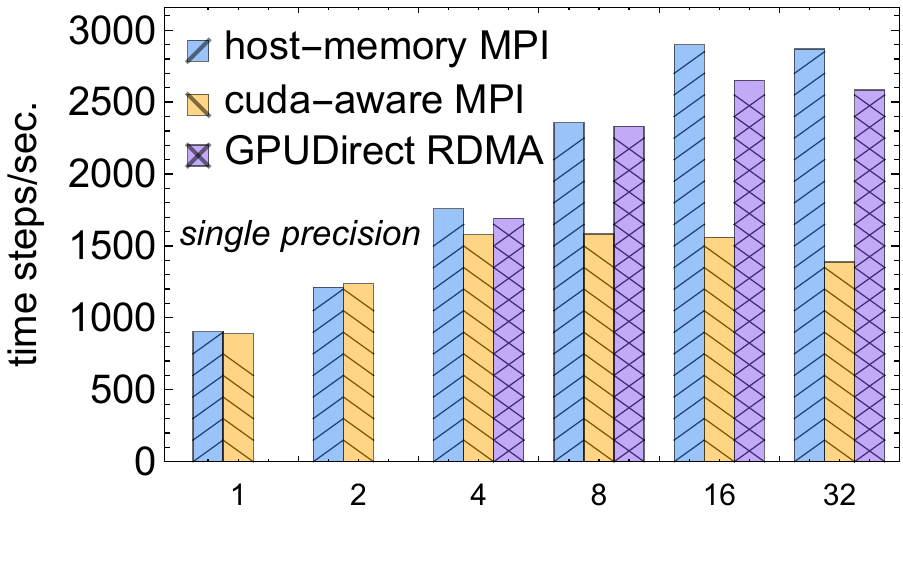}
\centering\includegraphics[width=\columnwidth]{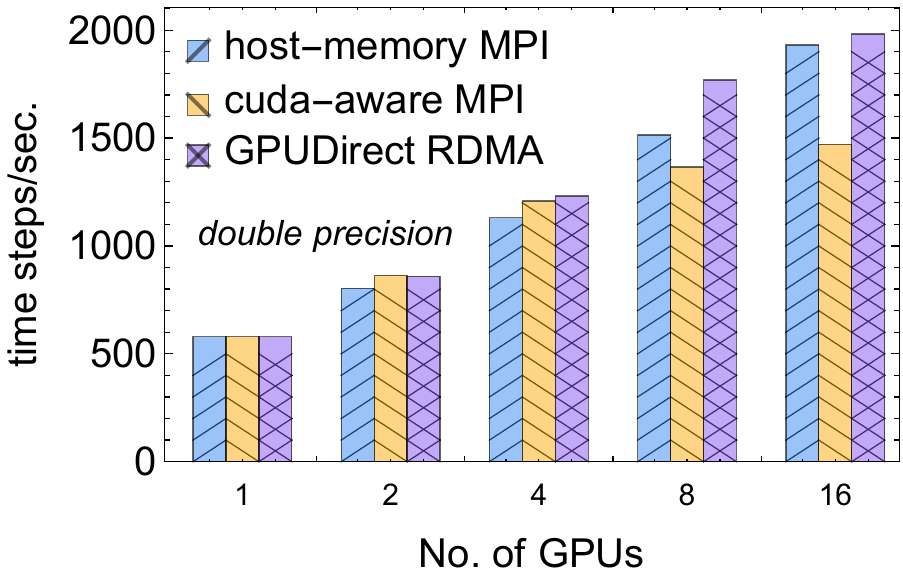}
\caption{(color online) Strong scaling benchmark for a GPUDirect RDMA benchmark of $N=64,000$
particles LJ liquid benchmark on the Wilkes GPU cluster, comparing default
host-memory MPI, CUDA-aware MPI, and CUDA-aware MPI with GPUDirect RDMA (GDR)
using MVAPICH 2.0 GDR (experimental).  Shown is the performance in number of
time steps per second vs. number of GPUs, for single precision ({\em top}) and
double-precision ({\em bottom}) runs.}
\label{fig:wilkes_lj}
\end{figure}

The Wilkes cluster at the University of Cambridge is a large GPU cluster
designed to fully leverage GPUDirect RDMA.  The system features 128 nodes
with two NVIDIA Tesla K20c GPUs for each node, dual-rail Mellanox Connect-IB
Infiniband and NVIDIA GPUDirect RDMA software stack.  The cluster is mainly
being used for academic research.  We assess the performance of HOOMD-blue on
this system by running the standard LJ benchmark that comes with HOOMD-blue,
with a total system size of $N=64,000$ particles (NVT, $\delta t=0.005$,
$r_{\mathrm{buff}}=0.4$, $T=1.2$, $r_\mathrm{cut}=3.0$, $\epsilon=\sigma=1.0$).
Fig.~\ref{fig:wilkes_lj} shows the scaling behavior, both for single-precision
(upper panel) and double-precision (lower panel) builds.  As can be seen, in
single precision the default version of HOOMD-blue, which does not take
advantage of CUDA-aware MPI at all, always performs better than either
CUDA-aware MPI or CUDA-aware MPI with GPUDirect RDMA.  The size of message
transferred in (32bit) single precision, which is half of that transferred in
(64bit) double precision, is too small to take advantage of full communication
bandwidth.

On the other hand, for double-precision mode, the lower panel of
Fig.~\ref{fig:wilkes_lj} shows a performance benefit of enabling CUDA-aware MPI
on up to 4 nodes, or $N/P=16,000$. More interestingly, we observe better
performance of GPUDirect RDMA than both plain CUDA-aware MPI and host memory
staging. This demonstrates the potential benefit of GPUDirect RDMA technology
in strong-scaling applications.

\begin{figure}
\centering\includegraphics[width=\columnwidth]{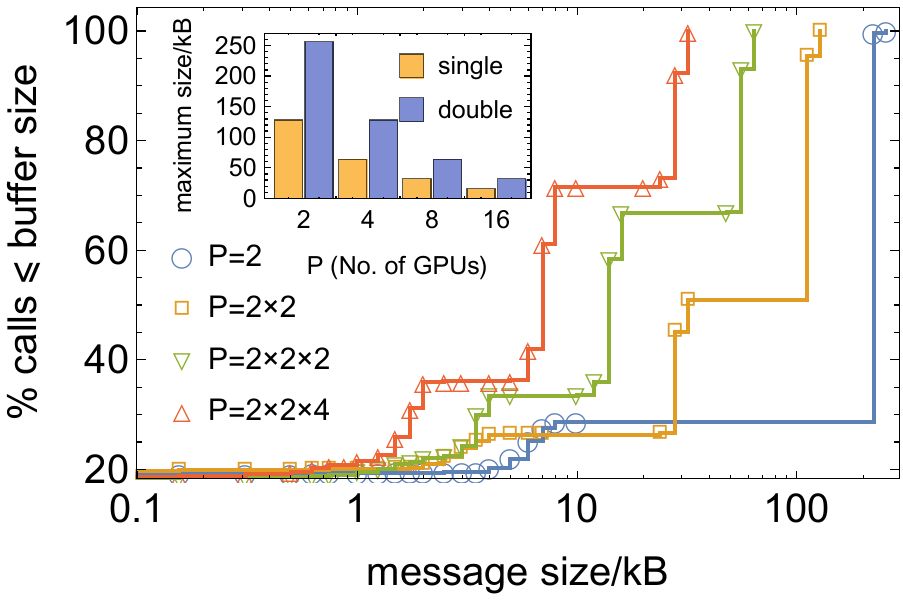}
%\centering\includegraphics[width=\columnwidth]{msg_size_avg_single_double}
\caption{(color online) Distribution of message sizes for the benchmark of
Fig.~\ref{fig:wilkes_lj} (lower panel).  Shown is the cumulative relative
frequency of message sizes in neighbor communication (MPI\_Isend/recv) in
double-precision runs, for different numbers of GPUs, obtained with the IPM
tool \cite{Skinner2005}.  {\em Inset:} Maximum message size in kB as function
of the number $P$ of GPUs, in single (light shaded/yellow) and double precision
(dark shaded/blue).}
\label{fig:message_size}
\end{figure}

Figure \ref{fig:message_size} shows the cumulative frequency of the
message size of {\tt MPI\_Irecv} calls during execution of the double-precision
LJ benchmark, for which data is shown in Fig.~\ref{fig:wilkes_lj} (bottom
panel).  The distribution has multiple 'knees', which are characteristic of the
communication pattern described in Sec.~\ref{sec:particle_migration}. The
precise location of these knees depends on the details of the domain
decomposition, however the maximum message size affects performance through
various internal thresholds of the MPI library. In the case of GPUDirect RDMA,
we were able to use maximum optimal thresholds of 32KB, above which the MPI
library switches to default pipelined communication. Interestingly, this limit
is reached with at least eight GPUs in single precision, or 16 GPUs in double
precision, for the $N=64,000$ LJ benchmark, as shown in
Fig.~\ref{fig:message_size}, inset. We confirm that for these minimum numbers
of GPUs the GPUDirect RDMA enabled benchmarks indeed perform superior to
cuda-aware MPI (Fig.~\ref{fig:wilkes_lj}), however the effective performance is
in the range of the optimized host memory implementation.

\subsection{Performance comparison to a CUDA-aware MPI enabled port of LAMMPS}
\begin{figure}
\centering\includegraphics[width=\columnwidth]{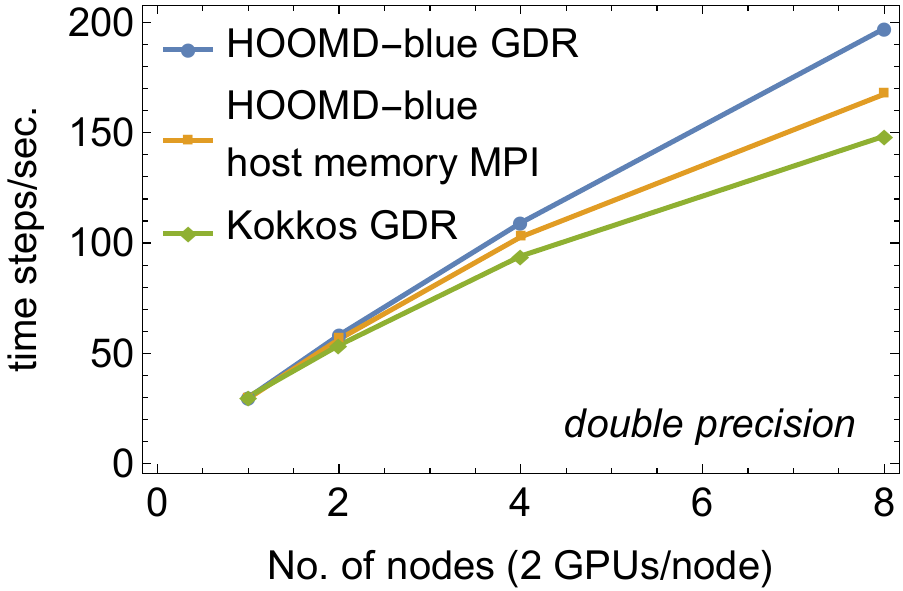}
\caption{(color online) Strong scaling of a double-precision LJ liquid benchmark of
$N=2,097,152$ particles comparing the performance of HOOMD-blue performance
(with GDR or with host memory MPI) and of LAMMPS-Kokkos (with GDR), on
$P=1\dots 8$ GPUs.  Shown is performance in time steps per second vs. number of
GPUs.}
\label{fig:comparison_lmp_kokkos}
\end{figure}

To assess whether HOOMD-blue makes optimal use of the CUDA-aware MPI based
communication protocols, we compare against another port of LAMMPS on GPUs,
LAMMPS-Kokkos, a recent alternative to LAMMPS-GPU.  The Kokkos package inside
LAMMPS is a forward looking capability with support for other accelerators
(Intel Xeon Phi), but with very limited feature support at the moment.  The
package supersedes a previous port of LAMMPS on GPUs, LAMMPS-CUDA
\cite{Trott2011}.  It also offers support for CUDA-aware MPI implementations,
which makes it interesting to compare to HOOMD-blue performance here.

As a benchmark system we choose the double precision LJ system benchmark
supplied with the Kokkos package (NVE, $N=2,097,152$, $\delta t=0.005$,
$r_{\mathrm{buff}}=0.3$, $r_\mathrm{cut}=2.5$, $\epsilon=\sigma=1.0$), with the
only change that we increase the neighbor list build frequency to every six
time steps, to ensure correct computation of forces. The corresponding
HOOMD-blue simulations start from the same fcc lattice initial configuration
(thermalized at $T=1.44$), additionally equilibrated over 30,000 time steps.
For the HOOMD-blue simulations we choose the optimal value of
$r_{\mathrm{buff}}$ and the distance check interval by prior tuning. Figure
\ref{fig:comparison_lmp_kokkos} shows the performance of the Kokkos package,
where LAMMPS is run in device communication mode and with GPUDirect RDMA
enabled, and for different build and runtime settings of HOOMD-blue. The
agreement of single-GPU performance emphasizes that both Kokkos and HOOMD-blue
are essentially fully optimized for simulations at this particle number, where
they are limited by device memory bandwidth. On the other hand, on eight nodes
the GDR version of HOOMD-blue performs better than the Kokkos package by a
factor of about 1.4, which we attribute to the optimizations of the
communication algorithm described in
Sec.~\ref{sec:gpu_communication_algorithm}.

\section{Strong scaling of a DPD benchmark}
\label{sec:strong_scaling_dpd}
\begin{figure}
\centering\includegraphics[width=\columnwidth]{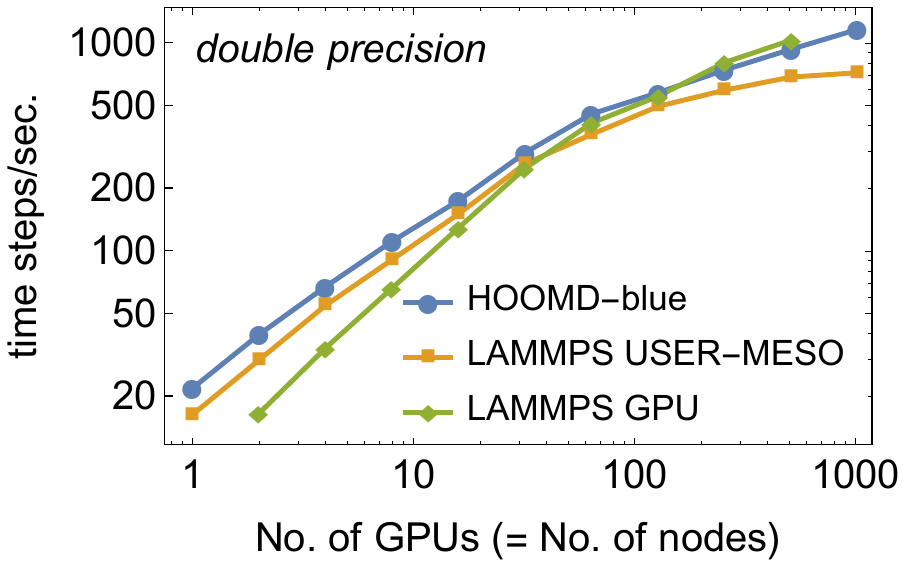}
\caption{(color online) Strong scaling performance of a DPD benchmark of $N=2,000,000$
particles, comparing HOOMD-blue, LAMMPS USER-MESO and LAMMPS-GPU. Shown is the
number of time steps per second, vs. number of GPUs, for $P=1\dots 1,024$
GPUs.}
\label{fig:comparison_lmp_dpd}
\end{figure}

We also compare dissipative particle dynamics (DPD) performance between
HOOMD-blue and two other codes. The communication pattern differs from that of
LJ in two ways. Velocities of ghost particles need to be communicated, in order
to compute the drag term in the DPD force. Hence, twice the amount of data is
communicated per time step. Moreover, to correctly seed per-particle-pair
random number generators \cite{Phillips2011}, global particle IDs of ghost
particles additionally need to be communicated with every ghost exchange.
Recently, Tang and Karniadakis \cite{Tang2014} presented a
GPU-optimized implementation of DPD, validated and benchmarked on the Titan
supercomputer. They demonstrate excellent strong scaling properties.  The
software is available within LAMMPS as the USER-MESO package
\cite{lammpsusermeso}.

Here, we compare HOOMD-blue performance on the Cray XK7 to the benchmark
numbers reported in Ref.~\cite{Tang2014} (Fig. 14 therein), for the same
benchmark of $N=2,000,000$ particles. We also compare to LAMMPS-GPU performance
\cite{Nguyen2014d}.  The simulation details are: $\delta t=0.005$ at number
density $\rho=3$, and with DPD parameters $A=25$, $\gamma=4.5$. HOOMD-blue and
LAMMPS-GPU use full double precision, the USER-MESO package is run in mixed
precision: communication is performed in double precision, but node local force
evaluations are performed in single precision.  As in the previous benchmarks,
HOOMD-blue performance was optimized by tuning the value $r_{\mathrm{buff}}$
and the distance check period. The number of time steps per second was measured
after an additional warm-up of 30,000 time steps, averaged over 50,000 steps.
LAMMPS-GPU and USER-MESO benchmarks were performed on Titan, HOOMD-blue
benchmarks were performed on the Blue Waters machine at the National Center for
Supercomputing Applications.  Even though this machine has the same
architecture as Titan (Cray XK7), we expect slight differences in performance,
from differences in system software and network configuration.  Figure
\ref{fig:comparison_lmp_dpd} shows the number of time steps per second as a
function of the number of GPUs in strong scaling, on up to 1,024 GPUs.
Performance values are maxima from several runs (up to 10\% variability was
observed for large runs).  Remarkably, the performance of HOOMD-blue parallels
that of the USER-MESO package over the whole range of numbers of GPUs and
appears to be only slightly superior (15\%), but outperforms LAMMPS-GPU for
small ($\lesssim 32$) numbers of GPUs.  However, for larger numbers of nodes
the performance between the codes is comparable, which we attribute to the
difference in communication patterns of DPD vs. LJ, where for DPD in double
precision, an amount of data four times larger is communicated than for LJ in
single precision. The DPD benchmark is therefore less sensitive to the latency
optimizations we focus on in this contribution, and given that it is likely
bandwidth-bound, further underscores that with GPUs the communication bandwidth
of current system architectures, along with various sources of latency, has
become the main limiting factor of MD performance.

\section{Conclusion and Outlook}
\label{sec:conclusion}
We gave a detailed account of how we ported HOOMD-blue to a distributed memory
model (MPI). Because HOOMD-blue is a fully GPU-enabled code, a particular
challenge was presented by the latency of device-to-device communication. We
addressed this challenge using a highly optimized communication algorithm.
Our communication routines are implemented on the GPU to reduce
the amount of data transferred over PCIe and allow us to take advantage
of CUDA-aware MPI libraries. We also optimized for strong scaling on thousands of GPUs, which
we achieved using a design for the neighbor list and force computation
kernels based on cooperative thread arrays and an auto-tuning algorithm.

We evaluate the performance of our code in terms of both weak and strong
scaling benchmarks, for which we compared it to similarly optimized
implementations of GPU-enabled MD, and find HOOMD-blue performance to be
equivalent or superior. HOOMD-blue exhibits qualitatively similar scaling
behavior to these other codes, indicating that our optimizations are
successful, and that the scaling limits inherent to the underlying architecture
have been reached.  We note that the GPU-centric design of HOOMD-blue is
different from other more traditional MD codes, which have started as CPU-only
codes.

In the case of GPUDirect RDMA, we find superior performance in double precision
benchmarks, demonstrating the usefulness of the technology, especially in
strong scaling situations, however also its current limitations.  To further
improve strong scaling performance, latency and bandwidth bottlenecks will have
to be reduced. Moreover, a closer integration of the GPU into the communication
path seems realistic, such as to provide the capability of GPU kernel
call-backs from MPI calls. In general, we anticipate that future designs will
tightly couple GPUs as throghput-optimized and CPUs as latency-optimized
compute components, and optimal code performance will depend on high-bandwidth
links between the processors \cite{nvlinkwhitepaper}, to achieve greater
concurrency.

In this first 1.0 release of HOOMD-blue with MPI, we did not enable multi-GPU
support for electrostatics calculations, rigid bodies or anisotropic particles,
available only in single-GPU simulations. We expect to implement these capabilities
in future versions. The current implementation exclusively relies on spatial
domain decomposition as a work distribution technique and thus applies
to mostly homogeneous systems, whereas sophisticated load-balancing
schemes have been implemented for more inhomogoeneous or biomolecular
systems \cite{Phillips2005,Hess2008,Grime2014} on CPUs, and they
should additionally benefit from GPU acceleration.

More broadly, our study further establishes GPUs as extremely fast engines for
MD simulation compared to traditional CPU cores. GPUs not only realize an order
of magnitude speed-up over current-generation CPUs, but they also scale very
well using spatial domain decomposition. Hence, our and comparable codes
therefore both greatly benefit from the unprecedented performance offered by
these fast processors, and at the same time they push the envelope of current
system designs.  Since the speed-ups presented here rely chiefly on exploiting
parallelism at various levels, i.e.\ by using GPUs on the node-level, and by
scaling the code up to many nodes, we provide a very clear case for how
parallelism can be the main enabling strategy in computational physics
discovery.

\paragraph{\bf Acknowledgments}
We gratefully acknowledge helpful discussions with Davide Rosetti, Yu-Hang Tang,
Dhabaleswar K. Panda, and Christian Trott. We thank Rong Shi for providing benchmark
data for GPUDirect RDMA in single precision.

This material is based upon work supported by the DOD/ASD (R\&E) under Award No.
N00244-09-1-0062 (JG, JAA, JAM, SCG). JG acknowledges support by DFG grant
GL733/1-1.  We also acknowledge support by the National Science Foundation,
Division of Materials Research, award DMR 1409620 (JAA and SCG), and award
DMR 0907338 (JG and DCM). This work was partially supported by a Simons
Investigator award from the Simons Foundation to Sharon Glotzer (SCG, JG).
This research used resources of the Oak Ridge Leadership Computing Facility at
the Oak Ridge National Laboratory, which is supported by the Office of Science
of the U.S. Department of Energy under Contract No. DE-AC05-00OR22725. This
research is part of the Blue Waters sustained-petascale computing project,
which is supported by the National Science Foundation (award number ACI
1238993) and the state of Illinois. Blue Waters is a joint effort of the
University of Illinois at Urbana-Champaign and its National Center for
Supercomputing Applications. We thank the University of Cambridge for providing
access to their Wilkes cluster.  Any opinions, findings, and conclusions or
recommendations expressed in this publication are those of the author(s) and do
not necessarily reflect the views of the DOD/ASD(R\&E). The Glotzer Group at
the University of Michigan is a CUDA Research Center. Hardware support by
NVIDIA is gratefully acknowledged.

\paragraph{Author contributions}
JG and JAA implemented the algorithm. TDN performed benchmarks on Titan, and JG
performed benchmarks on Blue Waters. PL and FS performed benchmarks on Wilkes.
JM provided the polymer brush benchmark.  DCM advised research at the
University of Minnesota and provided ideas for the implementation of the bond
communication algorithm, and SCG advised research at the University of
Michigan. All authors contributed to the manuscript.

%\biboptions{sort&compress}
%\bibliographystyle{cpc} \bibliography{md_gpu}

\begin{thebibliography}{10}

\bibitem{Anderson2008}
Anderson, J.~A., Lorenz, C.~D., and Travesset, A.,
\newblock Journal of Computational Physics {\bf 227} (2008) 5342.

\bibitem{Grand2012}
Grand, S.~L., G\"{o}tz, A.~W., and Walker, R.~C.,
\newblock Computer Physics Communications  (2012).

\bibitem{Plimpton1995}
Plimpton, S.,
\newblock J. Comp. Phys. {\bf 117} (1995) 1.

\bibitem{Trott2011}
Trott, C.~R.,
\newblock {\em {LAMMPSCUDA - a new GPU accelerated Molecular Dynamics
  Simulations Package and its Application to Ion-Conducting Glasses}},
\newblock PhD thesis, 2011.

\bibitem{Brown2011}
Brown, W.~M., Wang, P., Plimpton, S.~J., and Tharrington, A.~N.,
\newblock Computer Physics Communications {\bf 182} (2011) 898.

\bibitem{Tang2014}
Tang, Y.-H. and Karniadakis, G.~E.,
\newblock Computer Physics Communications {\bf 185} (2014) 2809.

\bibitem{Pall2013}
P\'{a}ll, S. and Hess, B.,
\newblock Computer Physics Communications {\bf 184} (2013) 2641.

\bibitem{Stone2007}
Stone, J.~E. et~al.,
\newblock Journal of Computational Chemistry {\bf 28} (2007) 2618.

\bibitem{Phillips2008}
Phillips, J., Stone, J., and Schulten, K.,
\newblock High Performance Computing, Networking, Storage and Analysis, 2008.
  SC 2008.  (2008) 1.

\bibitem{Eastman2013}
Eastman, P. et~al.,
\newblock Journal of Chemical Theory and Computation {\bf 9} (2013) 461.

\bibitem{Lysaght}
Lysaght, M., {Uchroński, Mariusz Kwiecien, Agnieszka Gebarowski, Marcin
  Nash}, P., Girottoa, I., and Todorovc, I.~T.,
\newblock PRACE whitepaper, PRACE-1IP .

\bibitem{Harvey2009}
Harvey, M.~J., Giupponi, G., and Fabritiis, G.~D.,
\newblock Journal of Chemical Theory and Computation {\bf 5} (2009) 1632.

\bibitem{Deng2011}
Deng, H., Li, X., Liu, X., and Wang, G.,
\newblock 2011 40th International Conference on Parallel Processing Workshops
  (2011) 191.

\bibitem{Roehm2012}
Roehm, D. and Arnold, A.,
\newblock The European Physical Journal Special Topics {\bf 210} (2012) 89.

\bibitem{Ganesan2011a}
Ganesan, N., Taufer, M., Bauer, B., and Patel, S.,
\newblock 2011 IEEE International Symposium on Parallel and Distributed
  Processing Workshops and Phd Forum  (2011) 472.

\bibitem{Colberg2011}
Colberg, P. and H\"{o}fling, F.,
\newblock Computer Physics Communications {\bf 182} (2011) 1120.

\bibitem{hoomd-url}
{HOOMD-blue, http://codeblue.umich.edu/hoomd-blue}.

\bibitem{CarterEdwards2014}
{Carter Edwards}, H., Trott, C.~R., and Sunderland, D.,
\newblock Journal of Parallel and Distributed Computing  (2014).

\bibitem{nvidiagpudirectrdma}
{GPUDirect RDMA, http://docs.nvidia.com/cuda/gpudirect-rdma/}.

\bibitem{Potluri2013}
Potluri, S., Hamidouche, K., Venkatesh, A., Bureddy, D., and Panda, D.~K.,
\newblock 2013 42nd International Conference on Parallel Processing  (2013) 80.

\bibitem{hoomdweb}
{HOOMD-blue homepage https://codeblue.umich.edu/hoomd-blue/publications.html}.

\bibitem{Glaser2014}
Glaser, J., Qin, J., Medapuram, P., and Morse, D.~C.,
\newblock Macromolecules  (2014) 140113062817000.

\bibitem{Glaser2014a}
Glaser, J., Medapuram, P., Beardsley, T.~M., Matsen, M.~W., and Morse, D.~C.,
\newblock Phys. Rev. Lett. {\bf 113} (2014) 068302.

\bibitem{Reith2011a}
Reith, D., Mirny, L., and Virnau, P.,
\newblock Progress of Theoretical Physics Supplement {\bf 191} (2011) 135.

\bibitem{Levine2011a}
Levine, B.~G. et~al.,
\newblock Journal of Chemical Theory and Computation {\bf 7} (2011) 4135.

\bibitem{Reith2011}
Reith, D., Milchev, A., Virnau, P., and Binder, K.,
\newblock EPL (Europhysics Letters) {\bf 95} (2011) 28003.

\bibitem{Lin2014a}
Lin, B., Martin, T.~B., and Jayaraman, A.,
\newblock ACS Macro Letters {\bf 3} (2014) 628.

\bibitem{Nguyen2014b}
Nguyen, T.~D., Carrillo, J.-M.~Y., Matheson, M.~a., and Brown, W.~M.,
\newblock Nanoscale {\bf 6} (2014) 3083.

\bibitem{Nguyen2014c}
Nguyen, T.~D., Fuentes-Cabrera, M., Fowlkes, J.~D., and Rack, P.~D.,
\newblock Physical Review E {\bf 89} (2014) 032403.

\bibitem{Lam2013}
Lam, C.-H. and Tsui, O. K.~C.,
\newblock Physical Review E {\bf 88} (2013) 042604.

\bibitem{Grime2014}
Grime, J. and Voth, G.,
\newblock Journal of Chemical Theory and Computation {\bf 10} (2014) 423.

\bibitem{Phillips2010}
Phillips, C.~L., Iacovella, C.~R., and Glotzer, S.~C.,
\newblock Soft Matter {\bf 6} (2010) 1693.

\bibitem{Wilms2012a}
Wilms, D., Virnau, P., Sengupta, S., and Binder, K.,
\newblock Physical Review E {\bf 85} (2012) 061406.

\bibitem{Zhang2013c}
Zhang, G., Stillinger, F.~H., and Torquato, S.,
\newblock Physical Review E {\bf 88} (2013) 042309.

\bibitem{Nguyen2012}
Nguyen, N. H.~P., Jankowski, E., and Glotzer, S.~C.,
\newblock Physical Review E {\bf 86} (2012) 011136.

\bibitem{Perlmutter2013}
Perlmutter, J.~D., Qiao, C., and Hagan, M.~F.,
\newblock eLife {\bf 2} (2013) e00632.

\bibitem{Benedetti2014}
Benedetti, F., Dorier, J., and Stasiak, A.,
\newblock Nucleic Acids Research {\bf 42} (2014) 10425.

\bibitem{Kapoor2013}
Kapoor, A. and Travesset, A.,
\newblock Proteins {\bf 81} (2013) 1200.

\bibitem{Martyna1994}
Martyna, G.~J., Tobias, D.~J., and Klein, M.~L.,
\newblock J. Chem. Phys. {\bf 101} (1994) 4177.

\bibitem{Martyna1996}
Martyna, G.~J., Tuckerman, M.~E., Tobias, D.~J., and Klein, M.~L.,
\newblock Molecular Physics {\bf 87} (1996) 1117.

\bibitem{Phillips2011}
Phillips, C.~L., Anderson, J.~A., and Glotzer, S.~C.,
\newblock Journal of Computational Physics {\bf 230} (2011) 7191.

\bibitem{Anderson2013a}
Anderson, J.~A. and Glotzer, S.~C.,
\newblock arxiv {\bf 1308.5587} (2013).

\bibitem{nvidiacudaawarempi}
{CUDA-aware MPI (nVidia),
  http://devblogs.nvidia.com/parallelforall/introduction-cuda-aware-mpi/}.

\bibitem{ModernGPU}
Baxter, S.,
\newblock {ModernGPU, http://nvlabs.github.io/moderngpu/}.

\bibitem{Bell2012}
Bell, N. and Hoberock, J.,
\newblock {Thrust: A Productivity-Oriented Library for CUDA},
\newblock in {\em GPU Computing Gems, Jade Edition}, edited by Hwu, W.-m.~W.,
  chapter~26, pages 359--373, Morgan Kaufmann, 2012.

\bibitem{Brown2012}
Brown, W.~M., Kohlmeyer, A., Plimpton, S.~J., and Tharrington, A.~N.,
\newblock Comp. Phys. Comm. {\bf 183} (2012) 449.

\bibitem{nvidiacudaprogrammingguide}
{CUDA Programming Guide, http://docs.nvidia.com/cuda/cuda-c-programming-guide}.

\bibitem{supplementary}
{Supplementary Information, at ...}

\bibitem{Paul1995}
Paul, W., Yoon, D.~Y., and Smith, G.~D.,
\newblock The Journal of Chemical Physics {\bf 103} (1995) 1702.

\bibitem{Millan:unpub}
Millan, J.,
\newblock {In preparation}.

\bibitem{nvidiagpudirect}
{GPUDirect Webpage, http://developer.nvidia.com/gpudirect}.

\bibitem{Shainer2011}
Shainer, G. et~al.,
\newblock Computer Science - Research and Development {\bf 26} (2011) 267.

\bibitem{Wang2011}
Wang, H. et~al.,
\newblock Computer Science - Research and Development {\bf 26} (2011) 257.

\bibitem{Wang2014b}
Wang, H., Potluri, S., Bureddy, D., Rosales, C., and Panda, D.~K.,
\newblock IEEE Transactions on Parallel and Distributed Systems {\bf 25} (2014)
  2595.

\bibitem{Skinner2005}
Skinner, D.,
\newblock {Performance monitoring of parallel scientific applications},
\newblock Technical report, Lawrence Berkeley National Laboratory (LBNL),
  Berkeley, CA, 2005.

\bibitem{lammpsusermeso}
{LAMMPS USER-MESO, http://www.cfm.brown.edu/repo/release/USER-MESO/}.

\bibitem{Nguyen2014d}
Nguyen, T.~D. and Plimpton, S.~J.,
\newblock Computational Materials Science {\bf in press} (2014).

\bibitem{nvlinkwhitepaper}
{NVLINK White paper, http://www.nvidia.com/object/nvlink.html}.

\bibitem{Phillips2005}
Phillips, J.~C. et~al.,
\newblock Journal of computational chemistry {\bf 26} (2005) 1781.

\bibitem{Hess2008}
Hess, B. and Kutzner, C.,
\newblock Journal of Chemical Theory and Computation {\bf 4} (2008) 435.

\end{thebibliography}

\end{document}